\documentclass[preprint,12pt]{elsarticle}
\usepackage[T1]{fontenc}
\usepackage[utf8]{inputenc}

\usepackage{amssymb}
\usepackage{amsmath}
\journal{New Astronomy}

\begin{document}

\begin{frontmatter}



\title{Cosmological constraints on viable $f(R)$ models using weak lensing}

\author[label1]{Leandro Pardo}
\ead{lmpardoc@unal.edu.co}

\author[label1]{Leonardo Castañeda}
\ead{lcastanedac@unal.edu.co}

\affiliation[label1]{
	organization={Universidad Nacional de Colombia, Sede Bogotá. Observatorio Astronómico Nacional (OAN). Grupo de Astronomía Galáctica, Gravitación y Cosmología. Bogotá},
	country={Colombia}.
}

%% Abstract
\begin{abstract}
The accelerated expansion of the universe represents one of the most important open problems in modern cosmology. Although the $\Lambda$CDM model provides an excellent description of a wide range of observations, the physical nature of dark energy remains unknown, motivating the exploration of alternative theories of gravity. Among these, $f(R)$ theories constitute a well-studied extension of General Relativity, capable of reproducing a $\Lambda$CDM-like background evolution without introducing an explicit dark energy component. However, such theories may lead to significant deviations at the level of structure growth, making them testable through observables sensitive to cosmological perturbations. In this work, we employ weak gravitational lensing as a probe to test several viable $f(R)$ gravity models. In particular, we analyze the impact of these theories on the matter power spectrum, as well as on the convergence and cosmic shear power spectra. The analysis is performed within a Bayesian framework using the Cobaya code, together with its modified gravity extension MGCobaya, which allows for the consistent generation of theoretical predictions and their confrontation with current observational data from weak lensing and CMB lensing measurements. Our results show that, for all the models considered, the standard cosmological parameters remain consistent with the $\Lambda$CDM scenario, as expected given that these models are constructed to mimic its background evolution. Nevertheless, we obtain non-trivial constraints on the characteristic parameters of several $f(R)$ models, exhibiting a strong model dependence and, in some cases, bounds reaching the $99\%$ confidence level. The results show that the models analyzed remain compatible with current data and highlight the importance of exploring and contrasting different parametrizations within the framework of $f(R)$ theories using cosmological observations.
\end{abstract}

%% Keywords
\begin{keyword}
Weak Lensing, $f(R)$ theories, Cosmological Parameter Constraints, MCMC.



\end{keyword}

\end{frontmatter}



\section{Introduction}
\label{int}
%% Labels are used to cross-reference an item using \ref command.

At the end of the last century, observations of supernovae led to the conclusion that the expansion of the universe is accelerating \cite{riess1998observational, perlmutter1999measurements}. This gave rise to the $\Lambda$CDM ($\Lambda$-\textit{Cold Dark Matter}) model  as an explanation of the observations within General Relativity, but it also introduced a new component known as dark energy, whose nature remains unknown. These findings motivated the introduction of alternative theories of gravity beyond General Relativity, known as \textit{Modified Gravity Theories}, among which $f(R)$ \textit{theories} are included. One of the most well-known and extensively studied models within this class is the Hu–Sawicki model \cite{hu2007models}, which has, among others, the advantage to reproduce the $\Lambda$CDM background evolution without the need to introduce dark energy. Other models, which will be discussed later in the course of this paper, also mimic the $\Lambda$CDM background evolution. For this reason, it is important to test the observational viability of $f(R)$ models using cosmological data, in order to assess their consistency with General Relativity and the $\Lambda$CDM scenario. 

A valuable tool for testing gravity models is \textit{weak lensing}, which refers to the deflection of light by gravitational potentials. The resulting effect is a subtle distortion of a large ensemble of background galaxies.  This distortion, caused by the large-scale structure acting on background galaxies, is known as \textit{cosmic shear} \cite{bartelmann2001weak}. Although different $f(R)$ models mimic the background evolution of $\Lambda$CDM, the same does not hold for their predictions of cosmic shear, which is what makes it an important technique for testing different theories of gravity. 

For the purposes described above, we use the Cobaya code to perform a statistical analysis of the different models. We employ the modified version of the CAMB code, known as MGCobaya, to implement the modified gravity models and to generate the theoretical predictions required for their comparison with observational data. This paper is organized as follows. Section \ref{sec:fr} briefly discusses $f(R)$ theories and the models tested in this work. Section \ref{sec:wl} provides an introduction to weak lensing and the related observables. Section \ref{sec:num} describes the numerical implementation in the Cobaya and MGCobaya codes, whose results are presented in Section \ref{sec:results}. Finally, the paper concludes with a summary of the main results in Section \ref{sec:conc}. Regarding the notation, we adopt natural units  $c=1$, where $c$ is the speed of light. The used metric signature is $(-+++)$. Greek indices take the values $0$ to $3$, while Latin indices $1$ to $3$. Background quantities are denoted with an overbar whenever it is necessary to distinguish them from perturbed quantities.

\section{Modified gravity $f(R)$ theories}
\label{sec:fr}

%MIRAR DEFINICIONES, NOTACIÓN, TÉRMINOS, ETC. YA. Se arregló la notación

In $f(R)$ theories we can derive the gravitational field equation from the action 
% \cite{de2010f}

\begin{equation}\label{accion}
	S= \frac{1}{16 \pi G} \int d^4x \sqrt{-g}[f(R)+\mathcal{L}_M],
\end{equation}

where $f(R)$ is a generalized function of the Ricci scalar $R$ and $\mathcal{L}_M$ is the \textit{matter Lagrangian}. From (\ref{accion}) we can get the $f(R)$ field equation \cite{de2010f}

\begin{equation}\label{campofr}
	f_R G_{\mu \nu}=8 \pi G T_{\mu \nu} ^{(m)}+\frac{1}{2}g_{\mu \nu}(f-Rf_R)+\nabla_\mu \nabla_\nu f_R-g_{\mu \nu} \Box f_R,
\end{equation}

with $f \equiv f(R)$ , $f_R \equiv \frac{\partial f}{\partial R}$ (and  $f_{RR} \equiv \partial^2_R  f, f_R^{(n)} \equiv \partial^{(n)}_R f$), $\nabla_{\mu}$ is the \textit{covariant derivative} and \textit{D'Alembertian} is  $\Box \equiv \nabla^\mu \nabla_\mu$. Making $f(R)=R-2\Lambda$  \textit{Einstein-Hilbert} action with \textit{cosmological constant}, $\Lambda$, is recovered. 

If we assume a flat universe, described by the FLRW (\textit{Friedmann-Lemaitre-Robertson-Walker}) metric, the Ricci scalar for the background cosmology is \cite{dodelson2020modern}

\begin{equation}\label{ricci}
	\bar{R}= 6\left(\dot{H}+2H^2\right).
\end{equation}
%esto es para friedmann background? R\ es independiente del modelo de gravedad. YA, abajo

The last expression is independent of the theory of gravity. Here, $H \equiv \dot{a}/a$ is the  \textit{Hubble parameter} with $a(t)$ the \textit{scale factor} and $\dot{}  \equiv d/dt$, with $t$ the \textit{cosmic time}.

For weak lensing treatment it is convenient to work with the perturbed FLRW metric in the \textit{Newtonian gauge}

\begin{equation}\label{gauge_newtoniano}
	ds^2= a^2(\tau)\left\{-\left[1+2\Psi(\boldsymbol{x}, \tau)\right]d\tau^2+\left[1-2\Phi(\boldsymbol{x}, \tau)\right]\delta_{ij}dx^idx^j\right\},
\end{equation}

%LLENAR VACÍOS DESDE ECUACIÓN DE CAMPO A PERTURBADAS. YA. Solo necesito éstas, ya dicho. Más adelante deberán escribirse completas. 

where we have shifted to the \textit{conformal time} $d \tau \equiv dt/a$.  The potentials $\Psi(\boldsymbol{x}, \tau)$ and $\Phi(\boldsymbol{x}, \tau)$, are the \textit{scalar perturbations}. With this gauge applied in (\ref{accion}) we can obtain the $f(R)$ perturbation equations. Our interest in the resulting equations is limited to the parametrization used by MGCobaya. In \textit{Fourier space} we have \cite{alex2019mgcamb}

%Considering the \textit{sub-horizon approximation}, $aH<<k$ and the \textit{quasi-static approximation} $aH \sim \partial/\partial \tau$ we have. YA. LO DEJO PORQUE SERVIRÁ MÁS ADELANTE.

\begin{equation} \label{P1}
	k^2 \Psi= -4 \pi G \mu(a,k) a^2 [\bar{\rho} \Delta + 3(\bar{\rho}+\bar{P}) \sigma]
\end{equation}

\begin{equation} \label{P2}
	k^2[ \Phi- \gamma(a,k) \Psi ]= 12 \pi G \mu(a,k) a^2(\bar{\rho}+\bar{P}) \sigma
\end{equation}

with $\bar{\rho}$ and $\bar{P}$ the density and pressure, $\Delta$ is the \textit{comoving density perturbation} and $\sigma$ is the \textit{anisotropic stress}. The parameter $\gamma \equiv \Phi/\Psi$ is the \textit{gravitational slip} and it is \cite{geng2015matter}

%$\bar{\rho} \Delta = \bar{\rho} \delta +3(aH/k)(\bar{\rho}+\bar{P})v $ $v$ is the \textit{velocity field}. YA. SIRVE PARA MÁS ADELANTE

\begin{equation}\label{gamma}
	\gamma= \frac{1+2\frac{k^2}{a^2}\frac{\bar{f}_{RR}}{\bar{f}_{R}}}{1+4\frac{k^2}{a^2}\frac{\bar{f}_{RR}}{\bar{f}_{R}}}.
\end{equation}

On its side $\mu$ allows defining an effective Newton constant  $G_{eff}=G\mu(a,k)$ where

\begin{equation}\label{mu}
	\mu= \frac{1}{\bar{f}_{R}} \frac{1+4\frac{k^2}{a^2}\frac{\bar{f}_{RR}}{\bar{f}_{R}}}{1+3\frac{k^2}{a^2}\frac{\bar{f}_{RR}}{\bar{f}_{R}}}.
\end{equation}

By means of Eqs. (\ref{gamma}) and (\ref{mu}), (\ref{P1}) and (\ref{P2}) can be solved for specific $f(R)$ models. From this set of equations, it follows that the evolution of perturbations is model dependent.

\subsection{Some representative $f(R)$ models.}

We will present some examples among the most known $f(R)$ models.

\subsubsection{Hu-Sawicki model.}

It was introduced in \cite{hu2007models} and it is given by the expression

\begin{equation}\label{HS}
	f(R)= R-m^2 \frac{c_1 (R/m^2)^n}{c_2 (R/m^2)^n +1}.
\end{equation}

Here $m^2 \equiv H_0^2 \Omega_m$, $H \vert_0 \equiv H_0$ is the \textit{Hubble constant}, $\Omega_i \equiv \bar{\rho}_{i,0}/\rho_{cr}$ are \textit{density parameters} today  (with \textit{critical density} defined as $\rho_{cr} \equiv 3H_0^2/8 \pi G$), where $i \in \left\lbrace \Lambda, m, b, c \right\rbrace$ for cosmological constant, matter, baryons and CDM, respectively (here we have considered only the matter species relevant for the purposes of this article), and $c_1, c_2$ are dimensionless model parameters. To account with these free parameters allows adjusting the models to the different observations like the ones of Solar System, and, of course, the acceleration in the expansion of the universe. Following the $\Lambda$CDM evolution implies that

\begin{equation}
	\frac{c_1}{c_2} = 6 \frac{\Omega_{\Lambda}}{\Omega_m}
\end{equation}

and

\begin{equation}
	\frac{c_1}{c_2^2}= -\frac{1}{n} f_{R0} \left[3 \left(1+4 \frac{\Omega_{\Lambda}}{\Omega_m}\right) \right]
\end{equation}

where $f_{R0}=f_R \vert_0$ (for details of the last two expressions refers to \cite{hu2016testing, li2018galaxy}). Actually, $f_{R0}$ is more commonly used than other parameters. It is a negative value but is usually expressed as positive, so one finds $|f_{R0}|$ simply as $f_{R0}$.

\subsubsection{Starobinsky model.}

Originally proposed in \cite{starobinsky2007disappearing}, the Starobinsky model obeys the equation

\begin{equation}
	f(R)= R-\lambda R_C \left[1- \left(1+\frac{R^2}{R_C^2}\right)^{-n}\right].
\end{equation}

where $R_C$ represents the  \textit{characteristic curvature} and $\lambda$ is a dimensionless characteristic model parameter. Typical values for $\lambda$ are in the unit order \cite{geng2015matter, starobinsky2007disappearing}.

\subsubsection{Other models.}

Among the many other different models we will show some specific ones according to the purpose of this paper. We consider the following models \cite{brax2008f,brax2012modified, perez2018cosmological}:

%agregar las otras referencias dadas en artículo. YA. Agregué todas las referencias necesarias. 

\textit{Logarithmic f(R) model}

\begin{equation}
	f(R)=R-2\Lambda-\xi \ln \left( \frac{R}{R_*}\right)
\end{equation}

and

\begin{equation}
	f(R)=R-2\Lambda\left( \frac{R}{R_*}\right)^w,
\end{equation}

\begin{equation}
	f(R)=R-\frac{2\Lambda}{1+b\sqrt{\frac{\Lambda}{R}}}
\end{equation}

%otros modelos. YA. Se dice en la implementación numérica

where $0<w \ll 1$ and $R_*$ plays the analogue role of $R_c$ of Starobinsky model. The free parameters $\xi$, $w$ and $b$ are characteristic of each model; when they tend to $0$ we recover the $\Lambda$CDM model. The last two models do not have a known established name. We will call here \textit{small power} and \textit{square root}, just as a way to be able to refer to them.

Since $\Lambda$CDM is the original attempt to explain the acceleration in the
expansion of the universe, they say that $f(R)$ models that are close to its expansion history \textit{mimic} $\Lambda$CDM \cite{pogosian2008pattern, zhao2009searching}. All of the models referred to here are designed to mimic $\Lambda$CDM background evolution, while the appropriate range of values for the free parameters is taken. In that case we can make the approximation $R\approx \bar{R}$ and, subsequently

\begin{equation} \label{approx}
	R \approx 3 H_0^2(\Omega_m a^{-3}+4 \Omega_{\Lambda}),
\end{equation}

%MIRAR LA RELACIÓN CON APROXIMACIÓN DE ALTA CURVATURA, fR<<1. YA. SE DICE EN WEAK LENSING, NO NECESARIO AQUÍ.

for Ricci scalar. This approximation can be directly derived from (\ref{ricci}), since we are assuming the Hubble parameter near $\Lambda$CDM for matter and cosmological constant domain. The approximation (\ref{approx}) is commonly considered in the literature \cite{hu2007models,  hu2016testing, li2018galaxy, zhao2009searching,oyaizu2008nonlinear, he2012testing,  brax2013impact,  reverberi2019frevolution, arnold2019realistic}, especially when making software simulations.

\section{Weak lensing}
\label{sec:wl}

The weak lensing effect manifests itself as a change in the shape of a background of galaxies, due to the presence of gravitational potentials. This effect cannot be inferred from a single object, therefore, the presence of certain types of phenomena, commonly associated with gravitational lensing, such as multiple images, cannot be expected, but rather small changes in a large number of sources. The mentioned effect is the cosmic shear and we will present below, starting with the \textit{lens equation}

\begin{equation}\label{lente}
	\beta^i= \theta^i-\alpha^i(\theta^j)
\end{equation}

where $\theta$ is the angular position of the image,  $\beta$ the angular position of the source, and  $\alpha$ is the reduced deflection angle, which can be written in terms of a  \textit{lensing potential} \cite{kilbinger2015cosmology}

\begin{equation}\label{pot_lente}
	\alpha^i(\theta^j)= \frac{\partial}{\partial \theta^i}\phi (\theta^j).
\end{equation}

In turn, the lensing potential can be written as \cite{lewis2006weak,Boubekeur_2014}

\begin{equation}\label{potencial_lente}
	\phi(\theta^j)= \int_0^{\chi} d \chi^{\prime} \frac{\chi-\chi^{\prime}}{\chi \chi^{\prime}}  [\Phi(\chi^{\prime}\theta^j, \chi^{\prime})+\Psi(\chi^{\prime}\theta^j, \chi^{\prime})],
\end{equation}

where the \textit{comoving distance} as function of the \textit{redshift} $z$, $\chi(z)$ is defined as:

\begin{equation}
	\chi(z)= \int_0^z \frac{dz'}{H(z')}.
\end{equation}

The previous equation indicates that photons coming from a source placed at a comoving distance $\chi$ are deflected from their unperturbed path by the gravitational potentials.

Taking the Jacobian of (\ref{lente})

\begin{equation}
	A= \frac{\partial \beta^i}{\partial \theta^j},
\end{equation}

which can be split as

\begin{equation} \label{A}
	A =
	\left(
	\begin{array}{cccc}
		1-\kappa-\gamma_t & -\gamma_\times   \\
		-\gamma_\times &  1-\kappa+\gamma_t & \\       
	\end{array}
	\right),	
\end{equation}

where $\kappa$, $\gamma_t$ and  $\gamma_\times$ are the \textit
{convergence},  the \textit{tangential shear} and the \textit{cross component of the shear}, respectively. The last two terms are components of the \textit{shear}, $\gamma=\gamma_t+i\gamma_\times$. We can switch to angular scales $\boldsymbol{l}$ by means of a two-dimensional \textit{Fourier transform}

\begin{equation}
T(\boldsymbol{\ell}) =
\int_{\mathbb{R}^2} \mathrm{d}^2\theta \,
T(\boldsymbol{\theta})\,
e^{i \boldsymbol{\ell}\cdot \boldsymbol{\theta}}
\end{equation}

to get a relation between the convergence and the lensing potential

\begin{equation}\label{kappa}
	\kappa(\boldsymbol{\ell})= \frac{1}{2}l^2 \phi (\boldsymbol{\ell}).
\end{equation}

\subsection{Power spectra}

For weak lensing analysis it is convenient to define the \textit{convergence power spectrum} $P_{\kappa}(l)$

\begin{equation}\label{conv_spectrum}
	\langle\kappa(\boldsymbol{\ell}) \kappa^*(\boldsymbol{\ell}^{\prime})\rangle=(2\pi)^2 \delta_D^{(2)}(\boldsymbol{\ell}-\boldsymbol{\ell}^{\prime})P_{\kappa}(l).
\end{equation}

where $\delta_D^{(n)}$ is the \textit{n-dimensional Dirac delta function}. The convergence power spectrum can be related to the \textit{matter power spectrum} $P(k,z)$, which, in turn, is defined by the equation

%definir delta como perturbación. Depende arriba de rho. YA. Definido abajo.

\begin{equation}\label{espectro}
	\langle \delta_m (\boldsymbol{k})  \delta_m(\boldsymbol{k}^{\prime}) \rangle= (2\pi)^3 P(k,z)\delta_D^{(3)}(\boldsymbol{k}-\boldsymbol{k}^{\prime})
\end{equation}

in which $\delta_m \equiv \frac{\rho_m-\bar{\rho}_m}{\bar{\rho}_m}$ is the \textit{matter perturbation} . Explicitly, the relation between these power spectra is \cite{dodelson2020modern}

\begin{equation}\label{P_kappa}
	P_{\kappa}(l)= \left(\frac{3}{2}\Omega_{m,0} H_0^2 \right)^2 \int_0^{\chi_{m}} d\chi \frac{1}{\bar{f}_R^2}  \frac{g_L^2(\chi)}{a^2(\chi)}P(k, \chi),
\end{equation}

with

\begin{equation}
	g_L(\chi) \equiv \int_{\chi}^{\chi_{m}}d \chi' \left(1-\frac{\chi}{\chi'} \right) W(\chi'),
\end{equation}

where $W(\chi)$ is a galaxy distribution normalized as $\int_0^{\infty} d\chi W(\chi)=1$, which must be included since there is not a single source but rather multiple ones; $\chi_m$ is the maximum comoving distance of the source sample and where the \textit{Limber approximation} \cite{limber1953analysis}, $k \approx l/\chi$, has also been taken into account.

Actually, the convergence spectrum is not an observable itself, but it can be demonstrated \cite{schneider2006gravitational} that 

\begin{equation} \label{P_shear}
	P_\gamma(l)=P_{\kappa}(l).
\end{equation}

The \textit{shear power spectrum} $P_\gamma(l)$, defined from the shear $\gamma$ in the same way as the convergence spectrum, is just the cosmic shear effect. The integral in (\ref{P_kappa}) reveals the fact that in weak lensing we do not have a single source. Along the light trajectory, the sources become lenses, and their dark matter is added to their baryonic matter. 

Besides the shear power spectrum, one could define the correlation functions $<\gamma_t \gamma_t >$  and $<\gamma_\times \gamma_\times >$; however, it is more convenient to define the combinations 

\begin{equation}
	\xi_\pm(\theta)= <\gamma_t \gamma_t > \pm <\gamma_\times \gamma_\times >
\end{equation}

which are the quantities most commonly provided by cosmic shear measurements. They are the \textit{shear correlation functions} and their respective spectra can be related to the convergence spectrum as \cite{schneider2006gravitational}

\begin{equation}\label{ximas}
	\xi_+(\theta)= \int_0^\infty \frac{dl l}{2 \pi} J_0 (l \theta) P_\kappa(l),
\end{equation} 

\begin{equation}\label{ximenos}
	\xi_-(\theta)= \int_0^\infty \frac{dl l}{2 \pi} J_4 (l \theta) P_\kappa(l),
\end{equation} 

where $J_n$ is the n-th order \textit{Bessel function}. 

Although gravitational lensing was originally thought of as an effect acting on light coming from individual, discrete sources such as galaxies, photons from CMB (\textit{Cosmic Microwave Background}) can also be lensed. This effect is called \textit{CMB lensing} and in that case it is preferred to define the \textit{lensing potential power spectrum} through \cite{dodelson2020modern}

\begin{equation}\label{lens_spectrum}
	\langle\phi(\boldsymbol{\ell}) \phi^*(\boldsymbol{\ell}^{\prime})\rangle=(2\pi)^2 \delta_D^{(2)}(\boldsymbol{\ell}-\boldsymbol{\ell}^{\prime})C_{\phi \phi}(l).
\end{equation}    

The effect of gravitational lensing impacts the CMB temperature and it allows obtaining information at higher redshifts. 

For the considered $f(R)$ models, taking into account that the proper parameters of the model ($f_{R_0}$, $w$, etc.) are $<<1$, we have that $f_R \approx 1$. It means that equation (\ref{P_kappa}) takes the known form for General Relativity.  Besides, since the background evolution is near $\Lambda$CDM , the comoving distance $\chi$ can be considered as the same for it and our $f(R)$ models. However, the matter power spectra $P(k,z)$ are not, because the predictions for the evolution of matter perturbations are different depending on the gravity model, as we pointed out in Section \ref{sec:fr}.

The previous discussion shows that depending on the gravity models the predictions for power matter spectra and, consequently, convergence, shear and shear correlation spectra are different, making weak lensing a powerful tool to test modified gravity theories. 

\section{Numerical implementation}
\label{sec:num}

We perform the statistical analysis for our $f(R)$ models through Cobaya (Code for Bayesian analysis in Cosmology)  \cite{torrado2021cobaya}. This code allows us to constrain cosmological parameters using Monte Carlo Markov Chain (MCMC) Metropolis-Hastings (MH) sampler \cite{lewis2002cosmological, lewis2013efficient}. In order to test the cosmological theories for our $f(R)$ models we use MGCobaya (Modified Growth with Cobaya) \cite{alex2019mgcamb, zhao2009searching, wang2023new,  hojjati2011testing,  lewis2000efficient} which is a new version of MGCAMB (Modified Growth with CAMB) that improves the compatibility with Cobaya. Let us detail how we perform the numerical implementations in these codes. 
%alguna introducción. ESTA DE ARRIBA.YA

\subsection{MGCobaya simulations}

MGCobaya (or MGCAMB) solves the perturbation equations (\ref{P1}) and (\ref{P2}). The so-called $\gamma$, $\mu$ parametrization \cite{alex2019mgcamb} implemented in MGCobaya allows testing different modified gravity models. For scalar-tensor theories \cite{faraoni2004scalar} the parameters take the expressions

\begin{equation}\label{gamma_general}
	\gamma= \frac{1- \frac{2k^2 \beta ^2}{k^2+m^2a^2}}{1+ \frac{2k^2 \beta ^2}{k^2+m^2a^2}}
\end{equation}

\begin{equation}\label{mu_general}
	\mu=1+ \frac{2k^2 \beta ^2}{k^2+m^2a^2}
\end{equation}

where $m$ is the \textit{scalaron mass}. Regarding $f(R)$ theories MGCobaya includes a built-in implementation of the Hu–Sawicki model for which

%confirmar/revisar lo de masa del escalarón. YA. CREO QUE PODRÍA DIFERIRSE A LA TESIS

\begin{equation}
	\beta=1/\sqrt{6},
\end{equation}

\begin{equation}\label{HS_m}
	m=m_0\left( \frac{4 \Omega_{\Lambda}+\Omega_m a^{-3}}{4 \Omega_{\Lambda}+\Omega_m }\right)^{(n+2)/2}
\end{equation}

and

\begin{equation}
	m_0= H_0 \sqrt{\frac{4 \Omega_{\Lambda}+\Omega_m}{(n+1)f_{R0}}}.
\end{equation}

More generally, the scalaron mass is given by \cite{pogosian2008pattern} 

\begin{equation}\label{scalaron}
	m= \sqrt{\frac{f_R}{3f_{RR}}}.
\end{equation}

After clarifying the different terms, note the consistency between equations (\ref{gamma}) and (\ref{mu}) and (\ref{gamma_general}) and (\ref{mu_general}). If we keep our analysis for models that mimic $\Lambda$CDM we can use the approximation (\ref{approx}). In that case we can calculate directly $\bar{m}(a)$ through

\begin{equation} \label{scalaron_approx}
	\bar{m}(a) \approx \sqrt{\frac{1}{3\bar{f}_{RR}}}.
\end{equation}

Actually, the last approximation is the one implemented by MGCobaya for Hu-Sawicki model (\ref{HS}), but we can modify it to apply to other models (besides, MGCobaya requires that $dm/dt$ expression to be also modified; the procedure is completely analogous). The modification is carried out in the \textbf{mgcamb.f90} module of MGCobaya. The parameter $f_{R0}$ is introduced into the modified equation to play the role of the characteristic parameter of each model: $w$, $b$, $\xi$  (this will also be applied in the Cobaya implementation discussed in the following subsection). The previous described method allows us to work with several $f(R)$. Indeed, it is enough to apply (\ref{scalaron_approx}) to the selected $f(R)$ model (different to Hu-Sawicki) together with the approximation (\ref{approx}). It is straightforward to modify the respective equations in MGCobaya. There are several viable functional forms that reproduce the $\Lambda$CDM background expansion like the \textit{Tsujikawa} \cite{tsujikawa2008observational} and \textit{exponential} \cite{zhang2006testing, cognola2008class, linder2009exponential, bamba2010cosmological} models, the ones proposed by Appleby–Battye \cite{appleby2007consistent, appleby2010curing} or those listed in the Perez and Nesseris paper \cite{perez2018cosmological} (of course, these examples do not exhaust all possible cases); but in this work we focus on a representative subset of models which worked in a stable way within the numerical implementation of MGCobaya.

%referenciar cobaya. YA

\subsection{Implementation in Cobaya}

We sample the standard parameters \textit{log amplitude of primordial scalar perturbations} $\log(10^{10} A_S)$, \textit{scalar spectral index} $n_S$, $H_0$,  $\Omega_b h^2$ and $\Omega_c h^2$ (with $h$ the \textit{dimensionless Hubble parameter}), besides, the characteristic free parameter  of the $f(R)$ models. We prefer to limit the number of cosmological parameters in order to ensure a stable and focused analysis, in which the signals arising from modified gravity are not masked by additional extensions of the cosmological model. According to this, we fix other parameters. For Starobinsky model we take $\lambda=1$, as it is known to be weakly constrained by current cosmological data and largely degenerate with the characteristic curvature scale. For Hu-Sawicki we fix $n=1$ and $R_C=R_*= 2 \Lambda$ for the models other than Hu–Sawicki. This is a natural choice, since requiring the recovery of the $\Lambda$CDM limit at high curvature implies $f(R)=R-2\Lambda$ \cite{starobinsky2007disappearing,bamba2010cosmological}.

We already mentioned that we performed the numerical calculation in MGCobaya, except for $\Lambda$CDM for which we use CAMB. About the likelihoods, we implement a set of them which are available in the Cobaya code. We will indicate them in the following. For local measurements of $H_0$ we select the  \textbf{H0.riess2020} \cite{riess2021cosmic}. This likelihood, together with \textbf{bao.desi 2024\_bao\_all}  \cite{adame2025desi} (BAO for \textit{Baryon Acoustic Oscillations}) allows breaking the degeneracies between background parameters $(H_0,\,\Omega_m)$. Among the likelihoods used, the most relevant for modified gravity is \textbf{DES Y1 shear} (Dark Energy Survey) \cite{abbott2018dark} which includes measurements from cosmic shear through the shear correlation functions (\ref{ximas}, \ref{ximenos}). Since weak lensing is directly sensitive to alterations in structure growth produced by $f(R)$, these modifications affect the matter power spectrum and this information is projected onto the convergence power spectrum. Via the equation (\ref{P_shear}), shear measurements constitute an essential observable for detecting possible deviations from $\Lambda$CDM in modified gravity theories. DES likelihood includes source galaxies until $z \lesssim 1.3$; its importance lies in the fact that deviations from $\Lambda$CDM are more noticeable at low redshifts \cite{de2010f, tsujikawa2007matter, tsujikawa2008effect}. In order to analyze the behavior for higher redshifts, ($z \approx 1\text{–}3$) we include \textbf{Planck 2018 lensing.CMBMarged} (CMB-marginalized, temperature+polarization-based lensing likelihood) \cite{aghanim2020planck, 2020}. It contains information about the CMB lensing potential power spectrum (\ref{lens_spectrum}), allowing to study the structure growth for higher redshift. It is a complement to the DES weak lensing analysis, helping to break growth–background degeneracies and extending the sensitivity to modified gravity effects at earlier cosmic times.  

In the Cobaya code we set the \textit{Gelman-Rubin criterion} \cite{gelman1992inference} with $R-1<0.1$. This is a typical value which ensures the convergence \cite{lemos2023robust, hernandez2019cosmological, krolewski2025measuring}. We apply nonlinear corrections at small scales using the HMCode version of Halofit \cite{mead2016accurate} available in MGCobaya. The prior values of the cosmological parameters are typical and they are preset by Cobaya. Regarding the priors for $f(R)$ models we select ranges keeping the parameter with small values since this is a condition for the models to not deviate from background $\Lambda$CDM evolution. To select those values we support in the spectra calculations in MGcobaya, such that they are enough to show a deviation of them with respect to $\Lambda$CDM (see examples in Section \ref{sec:results}). We summarize the prior ranges in Table \ref{tab:priors}. 

\begin{table}[t]
	\centering
	\begin{tabular}{lc}
		\hline\hline
		\textbf{Parameter} & \textbf{Prior} \\
		\hline
		$H_{0}$ & $[55,\, 91]$ \\
		$\Omega_{b}h^{2}$ & $[0.005,\, 0.1]$ \\
		$\Omega_{c}h^{2}$ & $[0.001,\, 0.99]$ \\
		$\log(10^{10}A_{\mathrm{s}})$ & $[1.61,\, 3.9]$ \\
		$n_{s}$ & $[0.8,\, 1.2]$ \\
		$f_{R0}$ & $[10^{-6},\, 10^{-3}]$ \\
		$\xi$ & $[10^{-12},\, 10^{-8}]$ \\
		$w$ & $[10^{-8},\, 10^{-4}]$ \\
		$b$ & $[10^{-5},\, 10^{-4}]$ \\
		\hline\hline
	\end{tabular}
	\caption{Priors adopted for the cosmological and $f(R)$ parameters.}
	\label{tab:priors}
\end{table}

\section{Results}
\label{sec:results}

\begin{figure}[!htbp]
	\centering
	\includegraphics[width=0.92\linewidth]{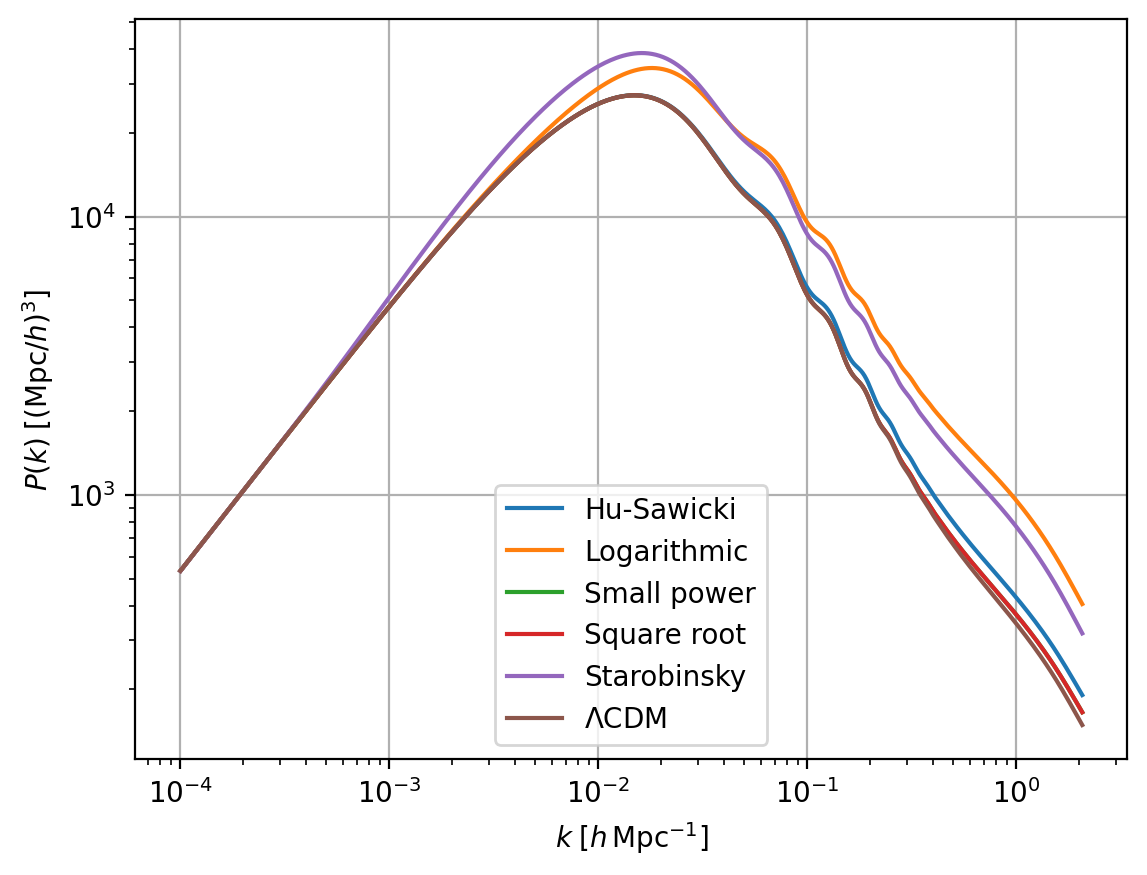}
\caption{Matter power spectra for $\Lambda$CDM and $f(R)$ models at $z=0$. We fixed values for $f(R)$ parameters according to the priors indicated in Table \ref{tab:priors}. We set  $f_{R0}=10^{-5}$, $\xi=10^{-10}$, $w=10^{-6}$ and $b=5 \cdot 10^{-5}$. All of the $f(R)$ models show an increase in the amplitudes of the spectra, especially at nonlinear scales.}\label{fig:matterpower}	
\end{figure}

After we detailed our numerical implementation, we can present the results of simulations for different $f(R)$ models. First, we include a set of power-spectrum plots associated with the likelihoods used in our analysis, as detailed in the previous Section. In Figure \ref{fig:matterpower} we show the matter power spectra for our tested $f(R)$. All of them present an enhancement with respect to $\Lambda$CDM, also shown in the plot, in the nonlinear regime. This behavior is well known and expected for $f(R)$ theories \cite{tsujikawa2008effect} and studied for several models like Hu-Sawicki \cite{peel2018breaking}, exponential and Starobinsky \cite{geng2015matter}.

\begin{figure}[!htbp]
	\centering
	\includegraphics[width=0.92\linewidth]{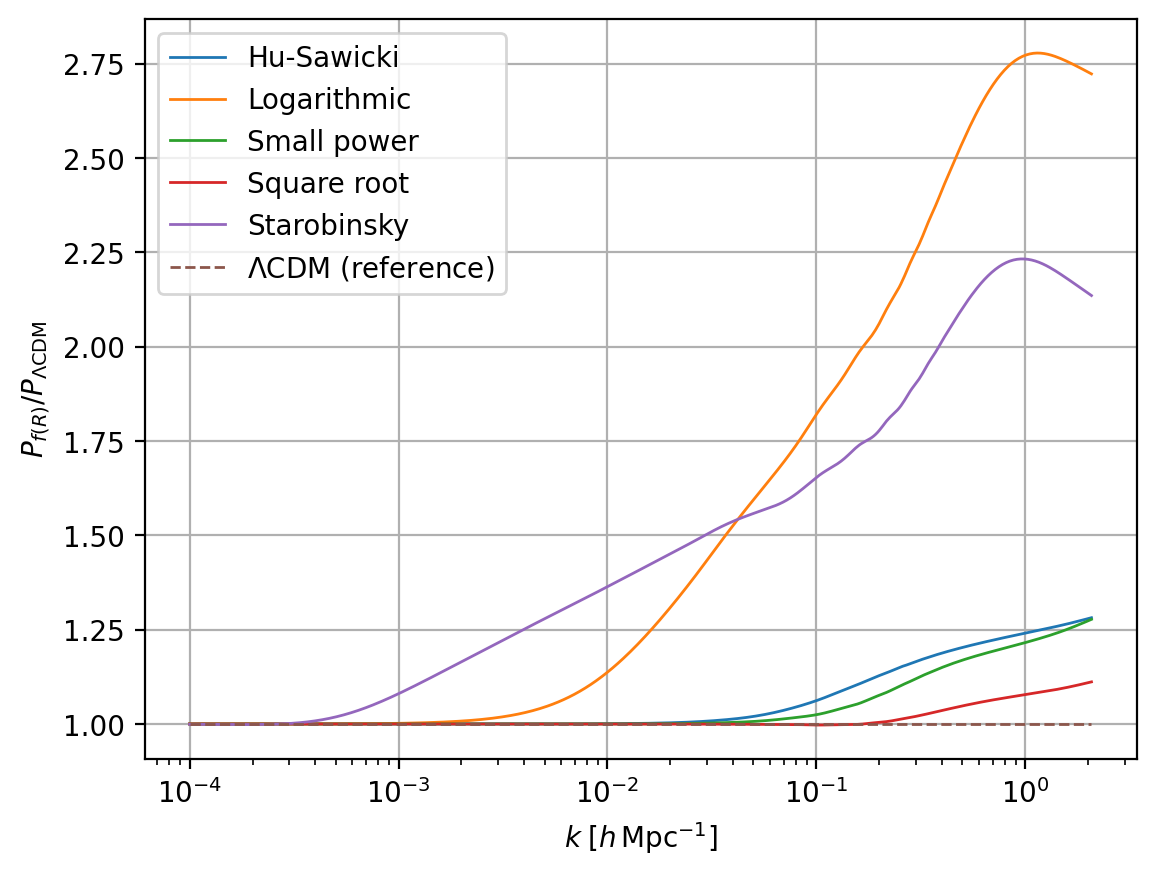}
\caption{Power matter spectra ratio to $\Lambda$CDM. The $f(R)$ models present deviations with respect to $\Lambda$CDM that range around from $10\%$ (small power) to $200\%$ (logarithmic). For Starobinsky and logarithmic deviations is notorious inclusive for scales from $k\sim 10^{-3}-10^{-4}$ $h Mpc^{-1}$.}\label{fig:relation_matterpower}	
\end{figure}

The mentioned behavior is clearer in Figure \ref{fig:relation_matterpower}, where we plot the ratios between the matter spectra of the $f(R)$ models with respect to the $\Lambda$CDM reference model We can observe the magnitude of it and the scales where it deviates from $\Lambda$CDM. Thus, square root model presents deviations, until around $10\%$, starting from $k\gtrsim 10^{-1} $ $hMpc^{-1}$ while for models like Hu-Sawicki and small power is more than double ($\sim 25-30\%$), from higher scales of $k\gtrsim 10^{-2}$ $hMpc^{-1}$. The enhancement is more noticeable for the remaining two models,  where the deviations reach more than  $100\%$ (Starobinsky), and almost $200\%$ (logarithmic) and they are significant from scales even much higher of  $k\sim 10^{-3}-10^{-4}$ $hMpc^{-1}$, respectively. We can conclude that the deviations from General Relativity are highly model-dependent, ranging from percent-level effects to order-unity enhancements.

\begin{figure}[!htbp]
	\centering
	\includegraphics[width=0.92\linewidth]{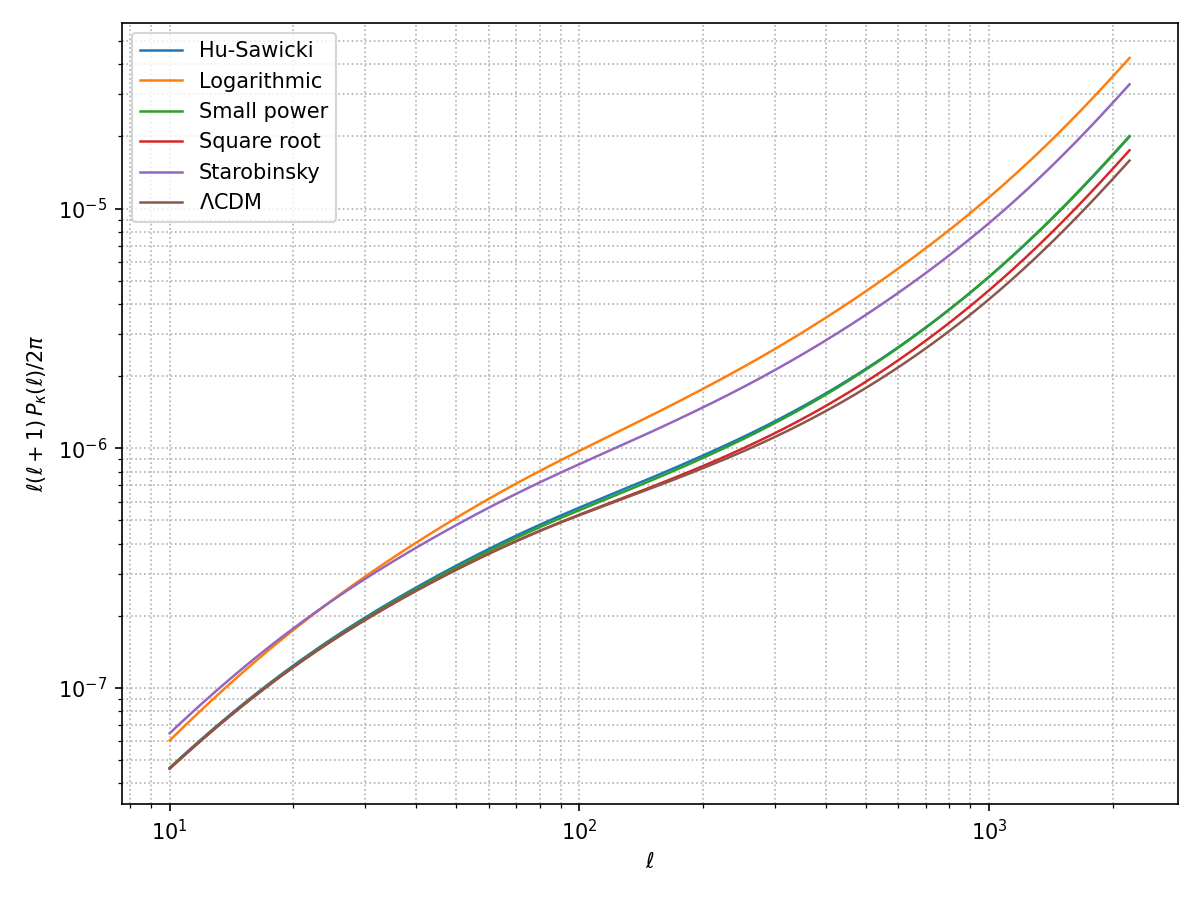}
\caption{Convergence power spectra for $\Lambda$CDM and $f(R)$ models. We took the same values for the parameters as in the matter power spectrum of Figure \ref{fig:matterpower}. Weak lensing amplifies the differences between $f(R)$ models and $\Lambda$CDM, making the departures especially for the logarithmic and Starobinsky cases, clearly visible across all multipoles.}\label{fig:convergencia}	
\end{figure}

The behaviour of matter power spectra is reflected in the convergence power spectrum. This is what is expected having in mind that it is a weighted integral of the matter power spectra according to (\ref{P_kappa}), and for our purposes is actually more important to analyze the convergence power spectra based on our discussion about weak lensing in Section $3$. In Figure \ref{fig:convergencia} we observe that logarithmic and Starobinsky models show the greatest deviation from $\Lambda$CDM for the whole range of multipolar scales. The other models are virtually indistinguishable at low multipoles but begin to differ from orders of magnitude of $\sim 10^2$. Comparable results have been obtained for Hu-Sawicki model, for example in \cite{peel2018breaking}. For all models, the deviation consists of an increase in the amplitudes of the convergence spectrum, which amplify as the multipolar scale increases. This behavior is consistent with the previously discussed observation that at small scales there was an enhancement in the matter power spectra. However, the convergence spectrum amplifies differences that were more modest in the matter power spectrum, showing that weak lensing is a particularly sensitive observable for discriminating modified gravity models. Thus, while there are no notable distinctions for linear scales in matter spectra with respect to $\Lambda$CDM, models such as logarithmic and Starobinsky show obvious differences for all multipolar scales.

\begin{figure}[!htbp]
	\centering
	\includegraphics[width=0.92\linewidth]{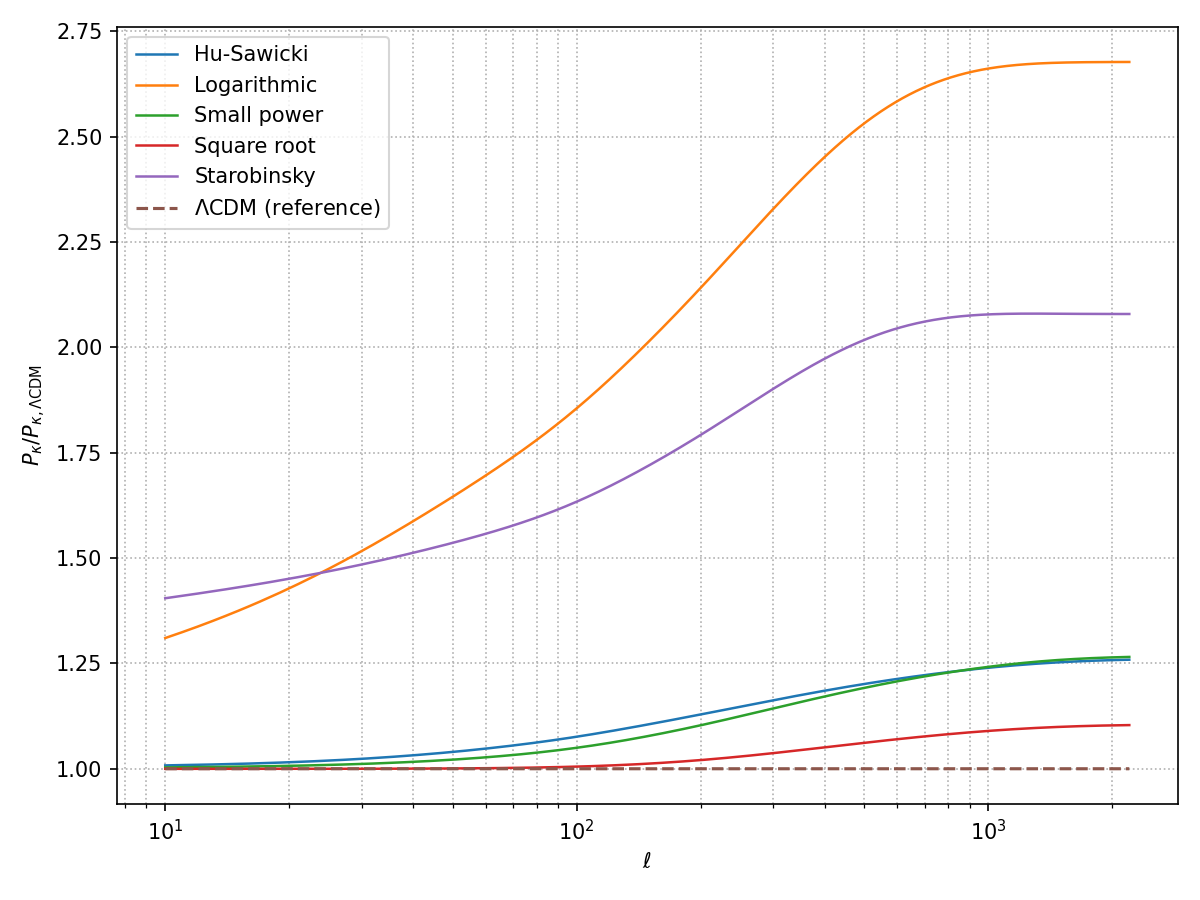}
\caption{Convergence spectra ratio to $\Lambda$CDM. The convergence spectra ratios make the hierarchy of departures from $\Lambda$CDM clearly visible, with large deviations for Starobinsky and logarithmic models and progressively milder ones for the remaining $f(R)$.} \label{fig:convergencia_ratio}	
\end{figure}

Similar to the analysis of the matter spectra, we also include plots of the convergence spectra ratio with respect to the $\Lambda$CDM reference model in Figure \ref{fig:convergencia_ratio}, which makes it easier to quantify the level of departures. The same hierarchy observed in matter spectra is maintained with similar levels. Starobinsky reaches until around $2.7$ times $\Lambda$CDM, logarithmic $2.1$, square root with moderate deviations, Hu-Sawicki, mild but increasing, both $\approx 1.25$, and small power is the closest to $\Lambda$CDM but even so, reaches until $\approx 1.1$. The differences in the power spectrum for small power are around $l \gtrsim 200 $, but for Hu-Sawicki and square root they are even from $l \gtrsim 50$. It is remarkable that differences for Starobinsky and logarithmic models are $\gtrsim 0.3$ above the $\Lambda$CDM prediction, even from low multipoles.

Although the effects that allow us to discriminate between different models already appear in the matter power spectra, they are more pronounced in the convergence spectra. In fact, the departures become evident at low multipoles, which, through the Limber approximation, correspond to large physical scales where the differences in the matter spectra remain mild. Weak lensing effectively amplifies the signatures imprinted in the matter power spectrum, as it integrates their contribution along the entire line of sight.

\begin{figure}[!htbp]
	\centering
	\includegraphics[width=0.92\linewidth]{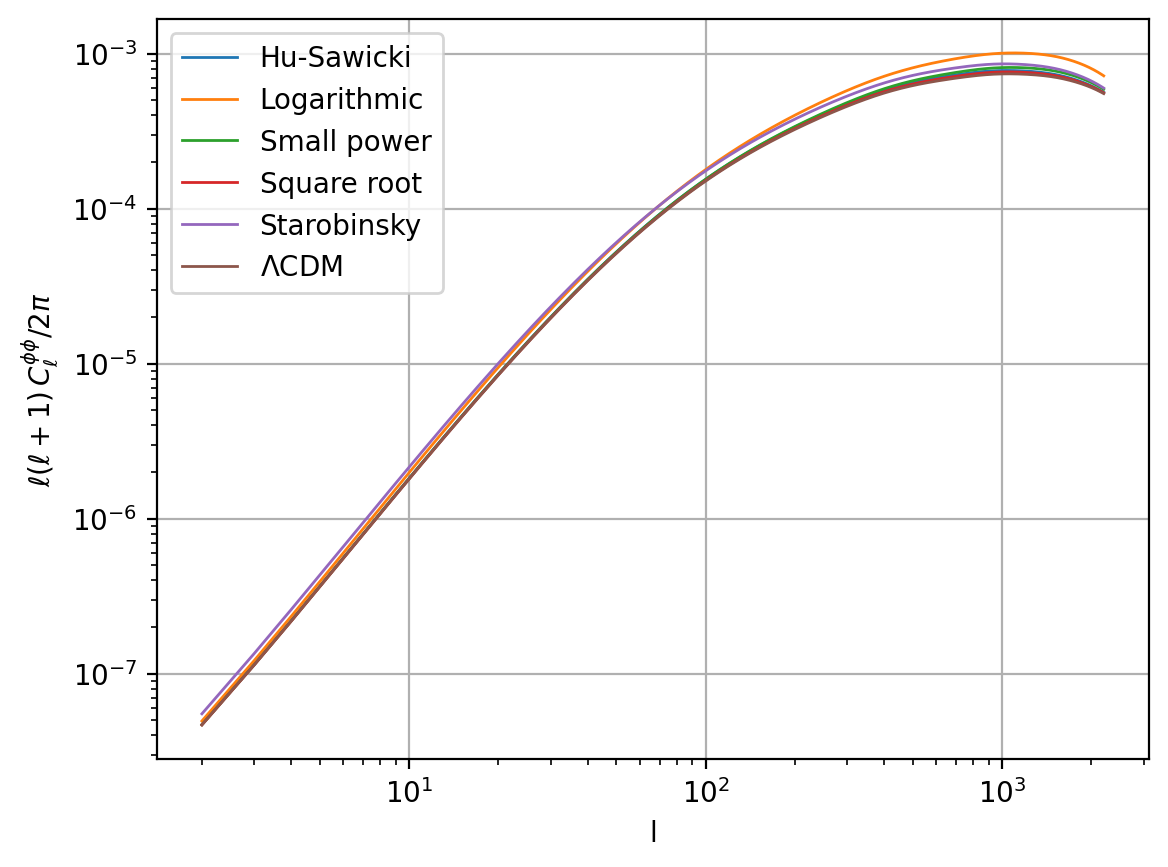}
\caption{CMB lensing potential power spectra for $\Lambda$CDM and $f(R)$ models. We took the same values for the parameters as in the matter and convergence power spectra (Figures \ref{fig:matterpower} and \ref{fig:relation_matterpower}, respectively). The behaviour obtained is consistent with that of the convergence spectra, in terms of the relative ordering of the models, the amplitudes of the lensing potential spectrum, and the multipole scales at which the effects appear.}\label{fig:clp}	
\end{figure}

\begin{figure}[!htbp]
	\centering
	\includegraphics[width=0.92\linewidth]{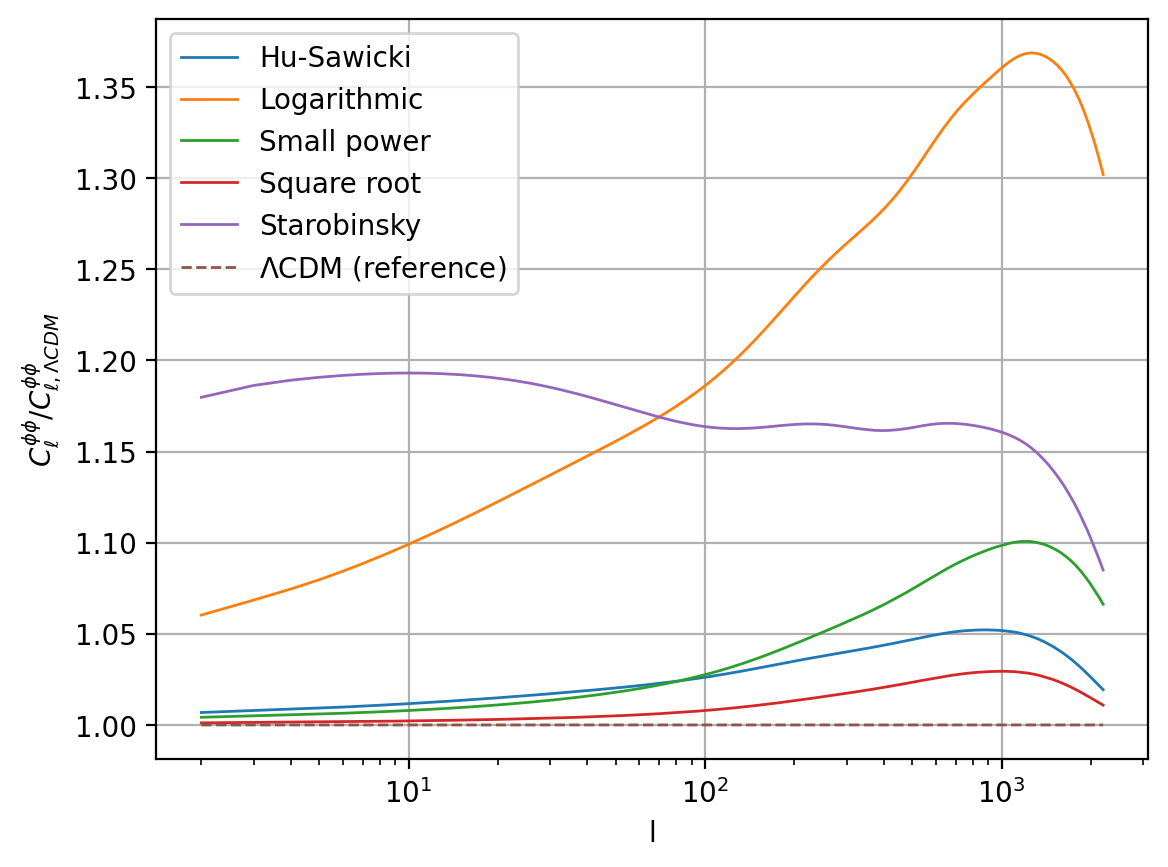}
\caption{CMB lensing potential spectra ratio to $\Lambda$CDM. Despite their smaller amplitudes, these spectra still offer meaningful constraints and serve as a valuable complement to weak lensing measurements.}\label{fig:clp_ratio}	
\end{figure}

In Figures \ref{fig:clp} and \ref{fig:clp_ratio} we plot the lensing potential power spectra and their ratios with respect to \textit{$\Lambda$}CDM, respectively, following the same structure used for the previous spectra. We observe that the same hierarchy among the different models is preserved. Thus, the results are consistent with those obtained for the convergence spectra, both in terms of the overall amplitudes and the multipole scales at which the effects appear. The amplitudes of these spectra lie in the range of about $3\%$ to $37\%$ above the $\Lambda$CDM reference. Although the departures are smaller than in the convergence spectrum, the CMB lensing potential spectrum still provides useful information and acts as a valuable complement to weak lensing observations. Actually, including the mentioned Planck  lensing likelihood noticeably improves the parameter inference, confirming its usefulness as a complementary probe. It should be recalled that the spectra shown correspond to specific choices of the $f(R)$ model parameters. However, these values are representative, since they lie within the prior ranges adopted in the parameter inference analysis and, importantly, those priors proved adequate: the MCMC chains converged and yielded well-defined constraints. Indeed, we now proceed to discuss the inference of the cosmological parameters obtained in our results.

\begin{figure}[!htbp]
	\centering
	\includegraphics[width=0.92\linewidth]{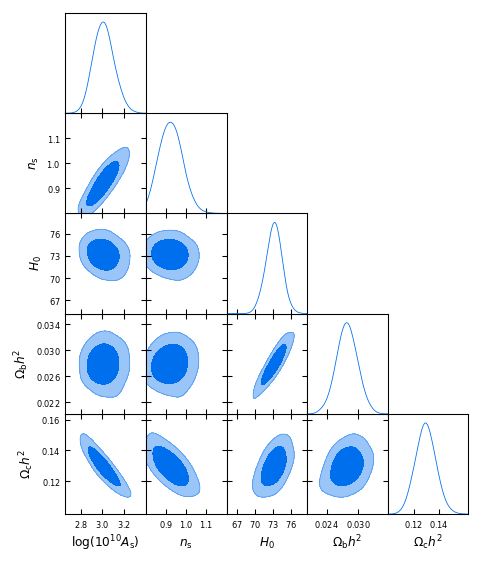}
\caption{Marginalized posterior distributions of the $\Lambda$CDM cosmological parameters. Results are consistent with the literature based on Cobaya framework \cite{hill2022atacama}.}\label{fig:rg_par}	
\end{figure}

Figure \ref{fig:rg_par} shows the marginalized posterior distributions of the $\Lambda$CDM cosmological parameters, presented as a triangle plot. Our results for $\Lambda$CDM model are consistent with the reported literature based on Cobaya framework (for example \cite{hill2022atacama}). This is very important since $\Lambda$CDM is our benchmark to validate our results for $f(R)$ models. This indicates that our selection of likelihoods is appropriate for constraining the cosmological parameters considered here. Summary of the marginalized cosmological constraints at $68\%$, $95\%$ and $99\%$ confidence levels (C.L.) for $\Lambda$CDM is shown in Table \ref{tab:constraints}.

\begin{figure}[t]	
\centering  
\includegraphics [scale=1.2]{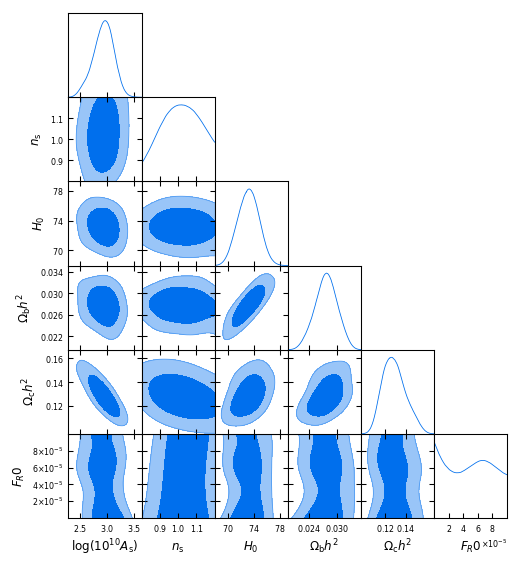}
\caption{Marginalized posterior distributions of the Hu-Sawicki cosmological parameters. We report a constraint of $f_{R0} < 6.63 \cdot 10^{-5}$ (68\% C.L.).}\label{fig:HS_par}	
\end{figure}

Now we proceed to present the results for the $f(R)$ models, where our main interest is to place constraints on the characteristic parameter of each model. We show the results for Hu-Sawicki model in Figure \ref{fig:HS_par}. The standard cosmological parameters ($\log \left( 10^{10}A_s \right)$, $n_s$, $H_0$, $\Omega_b h^2$, $\Omega_c h^2$) remain well constrained and take values compatible with $\Lambda$CDM. As shown in Table \ref{tab:constraints} we obtained

\begin{equation}
	f_{R0}<6.63 \cdot 10^{-5}   \quad (68\% \ \mathrm{C.L.}).
\end{equation}

\begin{figure}[!htbp]
	\centering
	\includegraphics[width=0.92\linewidth]{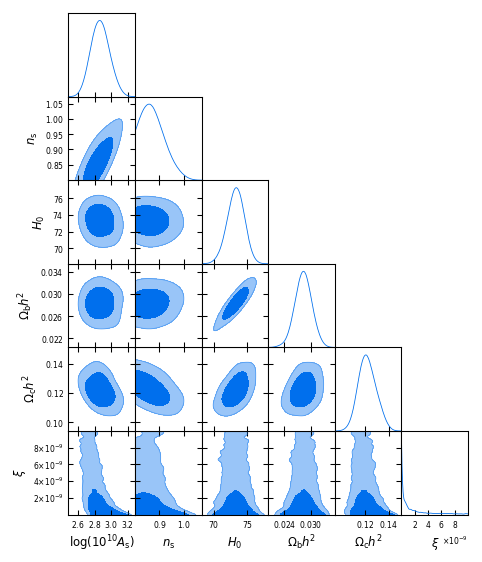}
\caption{Marginalized posterior distributions of the logarithmic cosmological parameters.  We obtained an upper bound constraint on the characteristic parameter $\xi$ at the $99\%$ C.L.}\label{fig:log_par}	
\end{figure}

The constraint corresponds to an upper bound at the $68\%$ confidence level; higher confidence levels are prior dominated and therefore not reported. It is consistent with previous results reported in the literature \cite{Boubekeur_2014}.

\begin{figure}[!htbp]
	\centering
	\includegraphics[width=0.92\linewidth]{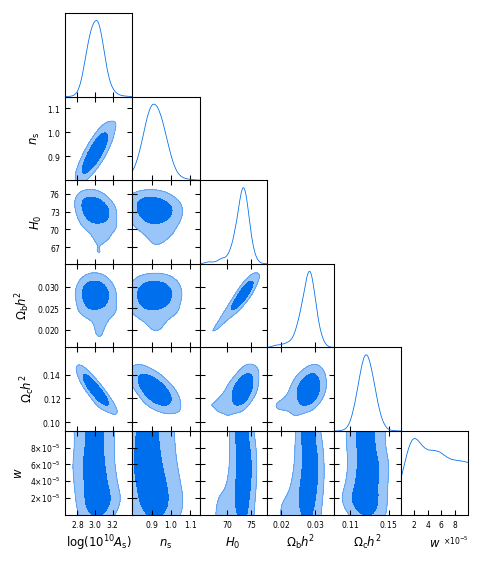}
\caption{Marginalized posterior distributions of the small power cosmological parameters. We obtained a constraint on the characteristic parameter $w$ with a well-defined maximum.}\label{fig:w_par}	
\end{figure}

\begin{figure}[!htbp]
	\centering
	\includegraphics[width=0.92\linewidth]{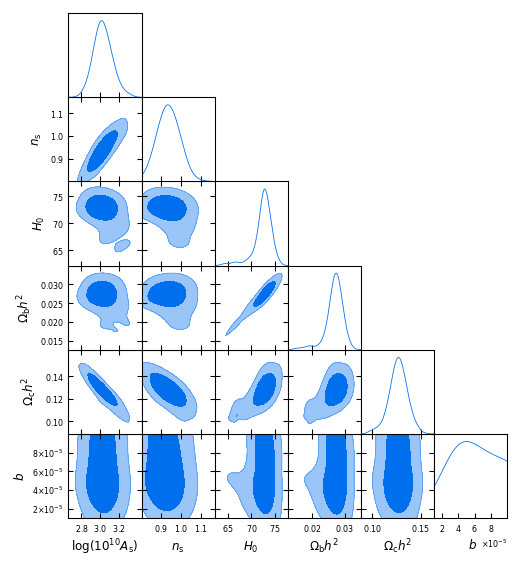}
\caption{Marginalized posterior distributions of the square root cosmological parameters. As in the small power case, this model also exhibits a well-defined maximum in the constraint on the characteristic parameter $b$.}\label{fig:sqrt_par}	
\end{figure}

\begin{figure}[!htbp]
	\centering
	\includegraphics[width=0.92\linewidth]{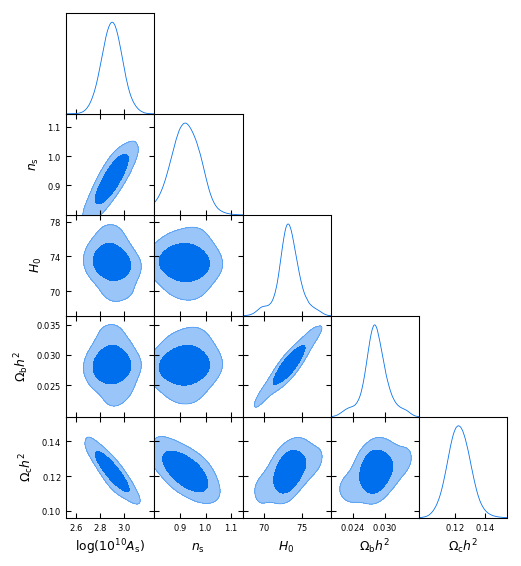}
\caption{Marginalized posterior distributions of the Starobinsky cosmological parameters. The standard cosmological parameters remain consistent with $\Lambda$CDM.}
\label{fig:str_par}	
\end{figure}

Regarding the other models, whose marginalized posterior distributions are shown in Figures \ref{fig:log_par}-\ref{fig:str_par}, the standard cosmological parameters remain well constrained too. We report the following constraints for the characteristic parameter for logarithmic, small power and square root models, respectively

\begin{equation}
	\xi<9.6 \cdot 10^{-9}   \quad (99\% \ \mathrm{C.L.}),
\end{equation}

\begin{equation}
	w=\left(4.8^{+2.2}_{-4.1}\right)\cdot10^{-5}   \quad (68\% \ \mathrm{C.L.}),
\end{equation}

\begin{equation}
	b=\left(5.7\pm2.4\right)\cdot10^{-5}   \quad (68\% \ \mathrm{C.L.}).
\end{equation}

For the small power and square root models, we obtain a well-defined maximum at the $68\%$ confidence level for their characteristic parameters. In addition, it is noteworthy that we can report constraints up to the $99\%$ confidence level for the characteristic parameter of the logarithmic model. The Starobinsky model is included for completeness; while its characteristic parameter is not constrained by the data, the standard cosmological parameters remain fully consistent with $\Lambda$CDM. All details regarding the constraints on the cosmological parameters and the characteristic parameters of each $f(R)$ model are shown in Table  \ref{tab:constraints}.

\begin{table}[t]
	\centering
	\begin{tabular}{lccc}
		\hline
		Parameter & 68\% & 95\% & 99\% \\
		\hline
		\multicolumn{4}{c}{\textbf{$\Lambda$CDM model}} \\
		\hline
		{\boldmath$\log(10^{10} A_\mathrm{s})$} & $3.012^{+0.089}_{-0.10}$ & $3.01^{+0.20}_{-0.18}$ & $3.01^{+0.26}_{-0.23}$ \\
		{\boldmath$n_\mathrm{s}$}               & $0.925^{+0.054}_{-0.063}$ & $0.92^{+0.11}_{-0.11}$ & $<1.07$ \\
		{\boldmath$H_0$}                        & $73.2\pm 1.4$ & $73.2^{+2.7}_{-2.9}$ & $73.2^{+3.6}_{-3.7}$ \\
		{\boldmath$\Omega_\mathrm{b}h^2$}       & $0.0278\pm0.0021$ & $0.0278^{+0.0041}_{-0.0041}$ & $0.0278^{+0.0053}_{-0.0053}$ \\
		{\boldmath$\Omega_\mathrm{c}h^2$}       & $0.1295\pm0.0084$ & $0.129^{+0.017}_{-0.016}$ & $0.129^{+0.023}_{-0.021}$ \\
		\hline
		\multicolumn{4}{c}{\textbf{Hu--Sawicki model}} \\
		\hline
		{\boldmath$\log(10^{10} A_\mathrm{s})$} & $2.93^{+0.21}_{-0.17}$ & $2.93^{+0.35}_{-0.39}$ & $2.93^{+0.42}_{-0.49}$ \\
		{\boldmath$n_\mathrm{s}$}               & $1.01^{+0.12}_{-0.10}$ & --- & --- \\
		{\boldmath$H_0$}                        & $73.2\pm1.6$ & $73.2^{+3.1}_{-3.1}$ & $73.2^{+4.0}_{-4.0}$ \\
		{\boldmath$\Omega_\mathrm{b}h^2$}       & $0.0277\pm0.0024$ & $0.0277^{+0.0045}_{-0.0050}$ & $0.0277^{+0.0052}_{-0.0055}$ \\
		{\boldmath$\Omega_\mathrm{c}h^2$}       & $0.1286^{+0.0093}_{-0.013}$ & $0.129^{+0.023}_{-0.020}$ & $0.129^{+0.029}_{-0.022}$ \\
		{\boldmath$F_{R0}$}                     & $<6.63\cdot10^{-5}$ & --- & --- \\
		\hline
		\multicolumn{4}{c}{\textbf{Logarithmic model}} \\
		\hline
		{\boldmath$\log(10^{10} A_\mathrm{s})$} & $2.87\pm0.11$ & $2.87^{+0.23}_{-0.21}$ & $2.87^{+0.29}_{-0.27}$ \\
		{\boldmath$n_\mathrm{s}$}               & $0.876^{+0.027}_{-0.067}$ & $<0.965$ & $<1.00$ \\
		{\boldmath$H_0$}                        & $73.3^{+1.3}_{-1.2}$ & $73.3^{+2.5}_{-2.6}$ & $73.3^{+3.0}_{-3.5}$ \\
		{\boldmath$\Omega_\mathrm{b}h^2$}       & $0.0283\pm0.0019$ & $0.0283^{+0.0037}_{-0.0037}$ & $0.0283^{+0.0048}_{-0.0055}$ \\
		{\boldmath$\Omega_\mathrm{c}h^2$}       & $0.1218^{+0.0071}_{-0.0084}$ & $0.122^{+0.016}_{-0.014}$ & $0.122^{+0.020}_{-0.019}$ \\
		{\boldmath$\xi$}                        & $<1.78\cdot10^{-9}$ & $<7.74\cdot10^{-9}$ & $<9.60\cdot10^{-9}$ \\
		\hline
		\multicolumn{4}{c}{\textbf{Small power model}} \\
		\hline
		{\boldmath$\log(10^{10} A_\mathrm{s})$} & $3.009\pm0.093$ & $3.01^{+0.18}_{-0.18}$ & $3.01^{+0.26}_{-0.22}$ \\
		{\boldmath$n_\mathrm{s}$}               & $0.918^{+0.051}_{-0.058}$ & $0.918^{+0.099}_{-0.11}$ & $<1.05$ \\
		{\boldmath$H_0$}                        & $72.9^{+1.8}_{-1.1}$ & $72.9^{+3.3}_{-3.9}$ & $72.9^{+3.8}_{-7.0}$ \\
		{\boldmath$\Omega_\mathrm{b}h^2$}       & $0.0277^{+0.0026}_{-0.0018}$ & $0.0277^{+0.0049}_{-0.0053}$ & $0.0277^{+0.0052}_{-0.0097}$ \\
		{\boldmath$\Omega_\mathrm{c}h^2$}       & $0.1268\pm0.0085$ & $0.127^{+0.017}_{-0.016}$ & $0.127^{+0.021}_{-0.021}$ \\
		{\boldmath$w$}                          & $\left(4.8^{+2.2}_{-4.1}\right)\cdot10^{-5}$ & --- & --- \\
		\hline
		\multicolumn{4}{c}{\textbf{Square root model}} \\
		\hline
		{\boldmath$\log(10^{10} A_\mathrm{s})$} & $3.032^{+0.095}_{-0.11}$ & $3.03^{+0.23}_{-0.20}$ & $3.03^{+0.30}_{-0.25}$ \\
		{\boldmath$n_\mathrm{s}$}               & $0.937\pm0.057$ & $0.94^{+0.11}_{-0.11}$ & $<1.07$ \\
		{\boldmath$H_0$}                        & $72.5^{+2.0}_{-1.0}$ & $72.5^{+3.8}_{-6.1}$ & $72.5^{+3.9}_{-8.5}$ \\
		{\boldmath$\Omega_\mathrm{b}h^2$}       & $0.0269^{+0.0027}_{-0.0017}$ & $0.0269^{+0.0055}_{-0.0079}$ & $0.0269^{+0.0056}_{-0.012}$ \\
		{\boldmath$\Omega_\mathrm{c}h^2$}       & $0.1268^{+0.0098}_{-0.0088}$ & $0.127^{+0.020}_{-0.020}$ & $0.127^{+0.023}_{-0.029}$ \\
		{\boldmath$b$}                          & $\left(5.7\pm2.4\right)\cdot10^{-5}$ & --- & --- \\
		\hline
		\multicolumn{4}{c}{\textbf{Starobinsky model}} \\
		\hline
		{\boldmath$\log(10^{10} A_\mathrm{s})$} & $2.898\pm0.089$ & $2.90^{+0.18}_{-0.18}$ & $2.90^{+0.23}_{-0.23}$ \\
		{\boldmath$n_\mathrm{s}$}               & $0.924\pm0.054$ & $0.92^{+0.10}_{-0.10}$ & $<1.05$ \\
		{\boldmath$H_0$}                        & $73.3\pm1.6$ & $73.3^{+3.5}_{-3.6}$ & $73.3^{+4.2}_{-4.4}$ \\
		{\boldmath$\Omega_\mathrm{b}h^2$}       & $0.0282\pm0.0024$ & $0.0282^{+0.0056}_{-0.0053}$ & $0.0282^{+0.0067}_{-0.0070}$ \\
		{\boldmath$\Omega_\mathrm{c}h^2$}       & $0.1226\pm0.0075$ & $0.123^{+0.015}_{-0.016}$ & $0.123^{+0.021}_{-0.018}$ \\
		\hline
	\end{tabular}
	\caption{Marginalized constraints on cosmological and model parameters at $68\%$, $95\%$, and $99\%$ confidence levels for $\Lambda$CDM and $f(R)$ models.}
	\label{tab:constraints}
\end{table}

\section{Conclusions}
\label{sec:conc}

In this paper we used weak lensing to test different $f(R)$ models. It was important to include weak lensing because it is directly sensitive to the growth of structure, and in $f(R)$ gravity the main deviations from $\Lambda$CDM manifest precisely as changes in the growth (and therefore in matter power spectra which are subsequently reflected in the convergence power spectra and in the shear correlation functions). According to our results, for all the considered 
$f(R)$ models, the standard cosmological parameters remain consistent with $\Lambda$CDM, indicating that the modified gravity extensions do not significantly affect the background cosmology, which is expected given that these $f(R)$ models are constructed to mimic $\Lambda$CDM background evolution.

For the tested models Hu-Sawicki, logarithmic, small power and square root, non-trivial constraints are obtained on their characteristic parameters, indicating that current data are sensitive to extensions of gravity beyond General Relativity. However, in none of the cases is the General Relativity limit excluded, and therefore none of the models can be conclusively ruled out. We could report constraints for the characteristic parameter, with well-defined maximum at $68\%$ C.L. for small power and square root models, and stringent upper limit at $99\%$ C.L. for the characteristic parameter of logarithmic model. In particular, the strong constraint obtained for the logarithmic model is not unexpected, given that it exhibits a large deviation from $\Lambda$CDM in the convergence power spectrum. These results highlight two important aspects. First, obtaining different outcomes for different models demonstrates the model dependence of the constraints, underscoring the importance of exploring parametrizations beyond the Hu–Sawicki model. Second, weak lensing, through the analysis of convergence power spectra, provides a particularly effective probe to discriminate among different realizations.

Previous studies within the Cobaya framework have explored parameter constraints for the Hu–Sawicki model \cite{mauland2023void, bai2025testing}. In addition, independent analyses have investigated the Tsujikawa model \cite{cen2019cosmological} using alternative MCMC implementations. In this work, we extend these efforts by incorporating a broader class of $f(R)$ gravity models within a unified Bayesian approach, allowing for a consistent and direct comparison of their phenomenology and observational viability under a common analysis strategy.

Regarding future work, the present analysis may be extended by considering additional viable $f(R)$ models, such as those discussed in Section~\ref{sec:fr}, enlarged parameter spaces, and a broader set of cosmological likelihoods.

%\appendix
%\section{Example Appendix Section}
%\label{app1}


%%       \cite{<label>} ==> [1]

%%
%Example citation, See \cite{lamport94}.




%\begin{thebibliography}{00}

%% For numbered reference style

%\bibitem{lamport94}
%  \textit{\LaTeX: a document preparation system},
 % Addison Wesley, Massachusetts,
 % 2nd edition,
 % 1994.

%\end{thebibliography}

\clearpage

\bibliographystyle{elsarticle-num}
\bibliography{biblio}

\end{document}